\documentclass[a4paper,10.5pt]{article}
\usepackage{geometry} 
\usepackage{graphicx,xcolor}
\usepackage{amsmath, amssymb, amsthm, bm, mathtools, amsfonts}
\usepackage{booktabs, multirow, tabularx, threeparttable}
\usepackage{comment, url, colortbl, setspace, ulem}
\usepackage{listings, algorithmic, algorithm, pb-diagram}
\usepackage[all]{xy}
\usepackage{ascmac, fancybox, latexsym}
\usepackage{autobreak}
\usepackage{bbm}

\theoremstyle{definition}

\definecolor{mygray}{rgb}{0.9,0.9,0.9}

\title{A Directional Measure of Marginal Inhomogeneity for Square Contingency Tables using Discrete-time hazard}

\author{Jun Tamura\thanks{Graduate School of Medicine, Yokohama City University, Japan}, \and
	Satoru Shinoda\thanks{Department of Biostatistics, School of Medicine, Yokohama City University, Japan}}

\begin{document}
	\maketitle
	\begin{abstract}
		In the analysis of square contingency tables with ordered categories, it is essential to assess deviations from marginal homogeneity (MH) when marginal equivalency between row and column variables does not hold. Some measures for evaluating the degree of departure from the MH model have been proposed. This study proposes a new directional measure using the discrete-time hazard, assuming that categories represent discrete time points. The proposed measure is capable of capturing both the magnitude and direction of deviation from the MH model. It is defined on a continuous scale from $-1$ to $1$, which allows for intuitive interpretation of the nature of marginal change.
		
		An estimator of the proposed measure and an asymptotic confidence interval are derived using the delta method. The theoretical properties of the measure are also discussed. The proposed measure provides a flexible tool for characterizing marginal inhomogeneity in square contingency tables under ordinal settings.
	\end{abstract}
	
	
	\vspace{0.5cm}
	
	\noindent\textbf{Correspondence:} \\
	Author Name: Jun Tamura \\
	Institution: Graduate School of Medicine, Yokohama City University, Japan \\
	Address: 3-9 Fukuura, Kanazawa-ku, Yokohama 236-0004, Japan\\
	Email: \texttt{light.130728ff@outlook.jp} \\

	\section{Introduction}
	
	In clinical research, it is essential to determine whether treatment effects improve or worsen over time compared to pre-intervention conditions. Traditional evaluation methods often rely on continuous variables, assessing effectiveness by calculating differences between pre- and post-intervention measures. However, continuous evaluation may not always be practical in real-world settings. For instance, Francom et al. \cite{francom1989} categorized sleep onset times into discrete ordinal categories before and after a two-week treatment period, comparing active sleep medication and placebo groups (Table \ref{tab:ta1}). This approach highlights the challenges of using continuous variables to assess treatment effects when responses are collected in discrete ordinal categories. In their study, subjects receiving the two treatments were treated as independent samples, simplifying the analysis but not accounting for potential within-subject variability. The primary objective was to evaluate whether sleep onset times improved or worsened post-intervention and to compare the effects between the two groups.
	
	These examples illustrate that, in clinical research, outcomes of interest are frequently observed in discrete ordinal categories \cite{gilboa2022, vaneekelen2020, huang2013}. The ordinal nature of such data poses significant challenges for applying conventional methods that rely on continuous variables, necessitating alternative approaches tailored to discrete data analysis.
	
	In square contingency tables where row and column variables share the same ordinal categories, such as Table \ref{tab:ta1}, observed values often cluster along the main diagonal. This clustering reflects the tendency of responses to remain consistent across pre- and post-intervention conditions, a common feature in ordinal data. This structure highlights the relevance of assessing the marginal distributions' equivalency rather than independence between rows and columns. The Marginal Homogeneity (MH) model, introduced by Stuart \cite{stuart1955}, is commonly used to evaluate the equivalency of marginal distributions. However, when the interest lies in the direction and magnitude of changes—such as pre- to post-intervention effects—quantifying deviations from marginal homogeneity becomes critical.
	
	Several measures for deviations from the MH model have been proposed \cite{tomizawa2003, tahata2006, tahata2008}. These measures primarily aim to quantify the extent to which marginal distributions retain uniformity or deviate from it. Additionally, to represent structures deviating from marginal equivalency in greater detail, Shinoda et al. \cite{shinoda2021} proposed the Marginal Continuation Odds Ratio (MCOR) model. This model uses marginal continuation odds to capture changes in distributions before and after intervention. It is particularly useful for evaluating the direction and magnitude of shifts observed pre- and post-intervention. Similarly, Ando et al. \cite{ando2024} proposed a measure of deviation from the MH model that corresponds to MCOR model, providing valuable insights into treatment effects. However, their measure is limited to a $0$--$1$ scale, which does not capture the direction of location shifts.
	
	To address this limitation, we propose a novel measure for evaluating deviations from marginal equivalency, designed to correspond to discrete-time hazards while also indicating the direction of shifts. 
	
	The remainder of this paper is structured as follows. Section 2 provides an overview of the basic models and scales for square contingency tables with ordinal categories. Section 3 introduces the proposed measure and its theoretical properties. Section 4 outlines the methodology for constructing asymptotic confidence intervals for the proposed measure. Section 5 present real-world data analysis to validate the methodology. Finally, the paper concludes with a summary of the findings and a discussion of potential applications and future research directions.

	\begin{table}[htbp]
		\centering
		\resizebox{\textwidth}{!}{%
			\begin{tabular}{p{1.5cm} p{1cm} p{1.5cm} p{1.5cm} p{1cm} p{0.5cm} p{0.5cm} p{1.5cm} p{1cm} p{1.5cm} p{1.5cm} p{1cm} p{0.5cm}}
				\multicolumn{6}{c}{Active Sleep Medication Group} & & \multicolumn{6}{c}{Placebo Group} \\
				\cmidrule(lr){1-6} \cmidrule(lr){8-13}
				Pre-intervention & \multicolumn{4}{c}{Post-intervention} & Total & & Pre-intervention & \multicolumn{4}{c}{Post-intervention} & Total \\
				\cmidrule(lr){2-5} \cmidrule(lr){9-12}
				& $<20$ & $20-30$ & $30-60$ & $>60$ & & & & $<20$ & $20-30$ & $30-60$ & $>60$ & \\
				\cmidrule(lr){1-6} \cmidrule(lr){8-13}
				$<20$ & 7 & 4 & 1 & 0 & 12 & & $<20$ & 7 & 4 & 2 & 1 & 14 \\
				$20-30$ & 11 & 5 & 2 & 2 & 20 & & $20-30$ & 14 & 5 & 1 & 0 & 20 \\
				$30-60$ & 13 & 23 & 3 & 1 & 40 & & $30-60$ & 6 & 9 & 18 & 2 & 35 \\
				$>60$ & 9 & 17 & 13 & 8 & 47 & & $>60$ & 4 & 11 & 14 & 22 & 51 \\
				\cmidrule(lr){1-6} \cmidrule(lr){8-13}
				Total & 40 & 49 & 19 & 11 & 119 & & Total & 31 & 29 & 35 & 25 & 120 \\
				\cmidrule(lr){1-6} \cmidrule(lr){8-13}
			\end{tabular}%
		}
		\caption{Comparison of Active Sleeping Medication and Placebo Groups}
		\label{tab:ta1}
	\end{table}

	\section{Contingency Table Analysis}
	
	In this section, we introduce models and measures important for analyzing \( r \times r \) square contingency tables. Section 2.1 explains the structure of the MH model and MCOR model using mathematical formulas. Section 2.2 describes the definition and properties of the measure proposed by Ando et al. \cite{ando2024}.
	
	\subsection{Model}
	
	\subsubsection{MH Model}

	Let the joint probability of the row variable \( X = i \) and the column variable \( Y = j \) be denoted as \( \Pr(X = i, Y = j) = p_{ij} \), where \( i = 1, \dots, r \) and \( j = 1, \dots, r \). The MH model can be expressed in various forms. For instance, the MH model can be defined in terms of cell probabilities as follows:
	\begin{equation*}
		p_{i\cdot} = p_{\cdot i}, \quad i = 1, \dots, r,
	\end{equation*}
	where \( p_{i\cdot} = \sum_{t=1}^r p_{it} \) and \( p_{\cdot i} = \sum_{s=1}^r p_{si} \). For further details on the MH model, refer to Stuart~\cite{stuart1955} and Bishop et al.~\cite{bishop1975}. The MH model indicates that the marginal distributions of the row and column variables are identical.
	
	The MH model admits several equivalent representations. In particular, when the categories are ordered, the formulation using continuation odds provides an insightful characterization:
	\begin{equation*}
		\omega_i^X = \omega_i^Y, \quad i = 1, \dots, r-1,
	\end{equation*}
	where
	\[
	\omega_i^X = \frac{p_{i\cdot}}{1 - F_{i-1}^X} = \Pr(X = i \mid X \geq i), \quad 
	\omega_i^Y = \frac{p_{\cdot i}}{1 - F_{i-1}^Y} = \Pr(Y = i \mid Y \geq i),
	\]
	and \( F_i^X = \sum_{k=1}^i p_{k \cdot} \), \( F_i^Y = \sum_{k=1}^i p_{\cdot k} \) denote the cumulative marginal probabilities for \( X \) and \( Y \), respectively.
	
	This equivalence also implies that the conditional probability of being in a lower category given survival beyond a higher one—i.e., \( \Pr(X \leq i \mid X \geq i+1) \) and \( \Pr(Y \leq i \mid Y \geq i+1) \)—is the same for both variables.

	In the context of square contingency tables where the categories represent discrete-time survival, the row variable's discrete-time survival probability can be expressed as:
	\begin{equation*}
		s_i^X = \sum_{k=i}^r p_{k\cdot} = \Pr(X \geq i) = 1 - F_{i-1}^X, \quad i = 1, \dots, r-1.
	\end{equation*}
	Thus, the row variable's discrete-time hazard function can be written as:
	\begin{equation*}
		1 - \frac{s_{i+1}^X}{s_i^X} = \frac{p_{i\cdot}}{s_i^X} = \Pr(X = i \mid X \geq i) = \omega_i^X, \quad i = 1, \dots, r-1.
	\end{equation*}
	
	Similarly, the column variable's discrete-time survival probability and hazard function can be defined analogously. Therefore, the discrete-time hazard \( \omega_i^X \) and \( \omega_i^Y \) can be interpreted as the discrete-time hazard functions for the row and column variables, respectively. 
	
	\subsubsection{MCOR model}
	In cases where the MH model does not hold, applying models of marginal inhomogeneity may be of interest. This is because, if such models fit the dataset well, the parameters of marginal inhomogeneity can quantify the degree of deviation from the MH model. Various models have been proposed to represent marginal inhomogeneity. In this section, we focus on models that are clinically interpretable within the context of discrete-time hazards. Specifically, we discuss the marginal continuation odds ratio (MCOR) model~\cite{shinoda2021}.
	
	The MCOR model is defined as follows:  
	\[
	\log\left(\frac{\omega_i^Y}{1-\omega_i^Y}\right) = \log\left(\frac{\omega_i^X}{1-\omega_i^X}\right) + \Delta, \quad i = 1, \ldots, r-1,
	\]  
	where the hazard of the column variable can be estimated as a positional shift of $\Delta$ in the logistic scale relative to the hazard of the row variable, where $\Delta$ represents the log odds ratio of the hazards. When $\Delta = 0$, the MCOR model reduces to the MH model. As $\Delta \to -\infty$ or $\infty$, the deviation from the MH model increases.

	\subsection{Measure}
	Various measures have been proposed to assess deviations from the MH model. Here, we introduce a recently proposed measure for assessing deviations from the MH model using discrete-time hazard. This measure is constructed based on the power divergence family, which is grounded in information-theoretic principles.
	Let \( W_{1(i)} = \omega_i^X(1-\omega_i^Y) \) and \( W_{2(i)} = \omega_i^Y(1-\omega_i^X) \) for \( i = 1, \ldots, r-1 \). For any real number \( \lambda > -1 \), Ando et al.~\cite{ando2024} proposed the following measure:
	\[
	\Psi^{(\lambda)} = \frac{\lambda(\lambda+1)}{2^\lambda - 1} I_{W}^{(\lambda)},
	\]
	where
	\[
	I_{W}^{(\lambda)} = \frac{1}{\lambda(\lambda+1)} \sum_{i=1}^{r-1} \left[ W_{1(i)}^* \left\{ \left( \frac{W_{1(i)}^*}{Q_{W(i)}^*} \right)^\lambda - 1 \right\} 
	+ W_{2(i)}^* \left\{ \left( \frac{W_{2(i)}^*}{Q_{W(i)}^*} \right)^\lambda - 1 \right\} \right],
	\]
	and
	\[
	Q_{W(i)}^* = \frac{1}{2}(W_{1(i)}^* + W_{2(i)}^*), \quad 
	W_{1(i)}^* = \frac{W_{1(i)}}{H_W}, \quad 
	W_{2(i)}^* = \frac{W_{2(i)}}{H_W}, \quad 
	H_W = \sum_{i=1}^{r-1} (W_{1(i)} + W_{2(i)}).
	\]
	
	This measure \( \Psi^{(\lambda)} \) has the following properties:
	\begin{itemize}
		\item \( 0 \leq \Psi^{(\lambda)} \leq 1 \),
		\item \( \Psi^{(\lambda)} = 0 \) if and only if the MH model holds,
		\item \( \Psi^{(\lambda)} = 1 \) if and only if the deviation from the MH model is maximized, corresponding to structures in the MCOR model where \( \Delta \to \infty \) or \( \Delta \to -\infty \).
	\end{itemize}
	
	While \( \Psi^{(\lambda)} \) is a highly practical and important measure, as it shares characteristics with existing indices such as Power-divergence and Diversity Index, it does not evaluate the direction of deviation. This limitation is evident in cases such as those shown in Table~\ref{tab:ta2}, where the measure yields the same value despite deviations occurring in opposite directions. This inability to distinguish between cases of improvement and deterioration during follow-up highlights its limitation in capturing directional changes.
	
	\begin{table}[htbp]
		\centering
		\begin{tabular}{cc}
			\renewcommand{\arraystretch}{0.6}
			\begin{tabular}{p{1.5cm}p{0.5cm}p{0.5cm}p{0.5cm}p{0.5cm}}
				\midrule
				$X$ & \multicolumn{3}{c}{$Y$} & \\
				\cmidrule(lr){2-4}
				& $1$ & $2$ & $3$ & Total \\
				\midrule
				$1$ & $0$ & $0$ & $0$ & $0$ \\
				$2$ & $0$ & $0$ & $0$ & $0$ \\
				$3$ & $0.4$ & $0.6$ & $0$ & $1$ \\
				\midrule
				Total & $0.4$ & $0.6$ & $0$ & $1$ \\
				\midrule
			\end{tabular}
			&
			\renewcommand{\arraystretch}{0.6}
			\begin{tabular}{p{1.5cm}p{0.5cm}p{0.5cm}p{0.5cm}p{0.5cm}}
				\midrule
				$X$ & \multicolumn{3}{c}{$Y$} & \\
				\cmidrule(lr){2-4}
				& $1$ & $2$ & $3$ & Total \\
				\midrule
				$1$ & $0$ & $0$ & $0.4$ & $0.4$ \\
				$2$ & $0$ & $0$ & $0.6$ & $0.6$ \\
				$3$ & $0$ & $0$ & $0$ & $0$ \\
				\midrule
				Total & $0$ & $0$ & $1$ & $1$ \\
				\midrule
			\end{tabular}
		\end{tabular}
		\caption{Comparison of Two Tables}
		\label{tab:ta2}
	\end{table}

	\section{Proposed Measure}
	In this section, we introduce a new measure to assess the direction and magnitude of deviations from the Marginal Homogeneity (MH) model using discrete-time hazard. Let \( W_{1(i)} = \omega_i^X(1-\omega_i^Y) \) and \( W_{2(i)} = \omega_i^Y(1-\omega_i^X) \). Assuming \( W_{1(i)} + W_{2(i)} \neq 0 \), we propose the following measure for marginal inhomogeneity:
	\begin{gather*}
		\Phi = \frac{4}{\pi} \sum_{i=1}^{r-1} (W_{1(i)}^* + W_{2(i)}^*)(\theta_i - \frac{\pi}{4}), \\
		W_{1(i)}^* = \frac{W_{1(i)}}{\Delta_{MH}}, \quad W_{2(i)}^* = \frac{W_{2(i)}}{\Delta_{MH}}, \quad \Delta_{MH} = \sum_{i=1}^{r-1}(W_{1(i)} + W_{2(i)}),
	\end{gather*}
	and
	\[
	\theta_i = \arccos\left(\frac{W_{1(i)}}{\sqrt{W_{1(i)}^2 + W_{2(i)}^2}}\right).
	\]
	
	The proposed measure \( \Phi \) has the following properties:
	\begin{itemize}
		\item \( -1 \leq \Phi \leq 1 \),
		\item \( \Phi = -1 \) if and only if \( W_{1(i)} = 0 \) and \( W_{2(i)} > 0 \) for all \( i \),
		\item \( \Phi = 1 \) if and only if \( W_{2(i)} = 0 \) and \( W_{1(i)} > 0 \) for all \( i \),
		\item \( \Phi = 0 \) indicates that the weighted sum of \( \theta_i - \frac{\pi}{4} \) is zero.
	\end{itemize}
	
	\subsection{Relationship Between the Proposed Measure and the MCOR Model}
	
	This section explains the relationship between the MCOR model, where $h^{-1}(\cdot)$ is defined as $\text{logit}(\cdot)$ and the proposed measure $\Phi$.
	
	Under the MCOR model, the proposed measure can be expressed as a function of $\Delta$:
	\[
	f(\Delta) := \frac{4}{\pi}\arccos\left(\frac{\exp(\Delta)}{\sqrt{\exp(2\Delta) + 1}}\right) - 1.
	\]
	\begin{figure}[htbp]
		\begin{center}
			\includegraphics[clip,width=15cm]{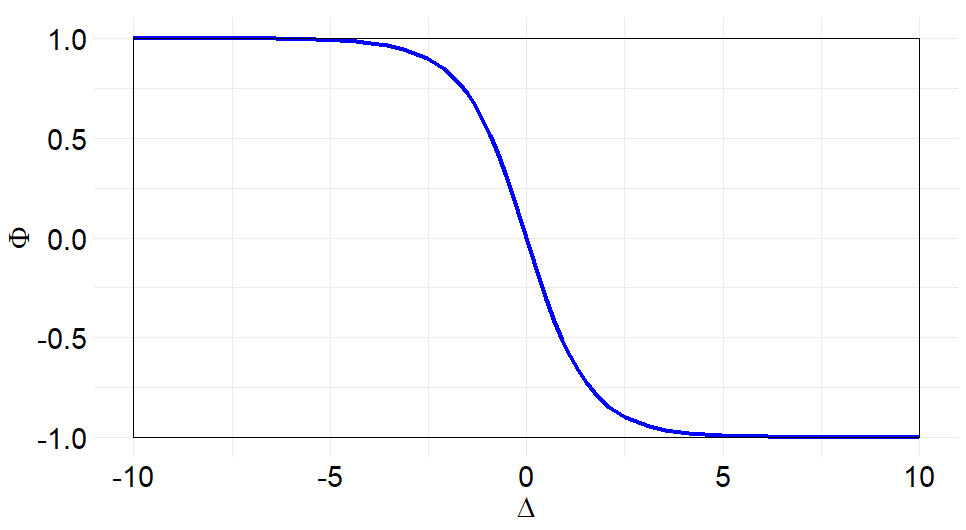}
			\caption{Changes in $\Phi$ with respect to $\Delta$ under the MCOR model.}
			\label{fig:figure1}
		\end{center}
	\end{figure}
	
	From the above, the relationship between the logistic-scale difference in discrete-time hazards $\Delta$ and $\Phi$ is illustrated in Figure \ref{fig:figure1}. The following properties hold under the MCOR model:
	\begin{itemize}
		\item The measure $\Phi$ is equal to 0 when the discrete-time hazards of the row and column variables are equivalent.
		\item As the discrete-time hazard of the row variable becomes larger than that of the column variable ($\Delta > 0$), the measure $\Phi$ approaches 1.
		\item Conversely, as the discrete-time hazard of the column variable becomes larger than that of the row variable ($\Delta < 0$), the measure $\Phi$ approaches -1.
	\end{itemize}

	\section{Approximate Confidence Interval for the Measure}

	This section outlines the construction of an asymptotic confidence interval for the proposed measure $\Phi$. The proposed measure $\Phi$ is expressed as:
	\[
	\Phi = \varphi(\bm{p}),
	\]
	where
	\[
	\bm{p} = (p_{11}, \ldots, p_{1r}, p_{21}, \ldots, p_{2r}, \ldots, p_{r1}, \ldots, p_{rr})^\mathsf{T},
	\]
	and $\mathsf{T}$ denotes the transpose operator. 
	
	Assuming that the observed frequencies $n_{ij}$ follow a multinomial distribution with cell probabilities $\bm{p}$ and total sample size $n = \sum_{i,j} n_{ij}$, the maximum likelihood estimator of $p_{ij}$ is given by $\hat{p}_{ij} = n_{ij}/n$. Thus, the estimator vector is $
	\hat{\bm{p}} = (\hat{p}_{11}, \ldots, \hat{p}_{1r}, \hat{p}_{21}, \ldots, \hat{p}_{2r}, \ldots, \hat{p}_{r1}, \ldots, \hat{p}_{rr})^\mathsf{T},
	$
	which is an $r^2 \times 1$ vector. The empirical estimate of the measure is then:
	\[
	\widehat{\Phi} = \varphi(\hat{\bm{p}}).
	\]
	
	By the central limit theorem, we have:
	\begin{gather*}
		\sqrt{n}(\hat{\bm{p}} - \bm{p}) \xrightarrow{d} N\left(\bm{0}_{r^2}, \xi(\bm{p})\right), \quad \text{as } n \rightarrow \infty, \\
		\xi(\bm{p}) = \mathrm{diag}(\bm{p}) - \bm{p}\bm{p}^\mathsf{T},
	\end{gather*}
	where $\bm{0}_{r^2}$ is the $r^2$-dimensional zero vector, and $\mathrm{diag}(\bm{p})$ denotes the $r^2 \times r^2$ diagonal matrix with entries of $\bm{p}$ on the diagonal.
	
	Applying the delta method, the distribution of $\varphi(\hat{\bm{p}})$ can be approximated as:
	\[
	\sqrt{n}(\varphi(\hat{\bm{p}}) - \varphi(\bm{p})) \xrightarrow{d} N\left(0, \nabla\varphi(\bm{p})^\mathsf{T} \, \xi(\bm{p}) \, \nabla\varphi(\bm{p})\right), \quad \text{as } n \rightarrow \infty,
	\]
	where $\nabla$ denotes gradient.
	
	For sufficiently large $n$, the $100(1 - \alpha)\%$ confidence interval for the measure $\Phi$ is approximated by:
	\[
	\varphi(\hat{\bm{p}}) \pm z_{\alpha/2} \sqrt{\frac{\nabla\varphi(\hat{\bm{p}})^\mathsf{T} \, \xi(\hat{\bm{p}}) \, \nabla\varphi(\hat{\bm{p}})}{n}},
	\]
	where $z_{\alpha/2}$ is the $(1 - \alpha/2)$ quantile of the standard normal distribution.
	
	\section{Real data analysis}
	We applied the proposed measure $\Phi$ to the data presented in Table~\ref{tab:ta1}. The analysis is based on a randomized controlled trial reported by Francom et al.\cite{francom1989}, which investigated whether sleep latency improved or worsened after intervention, and aimed to compare the effects between two groups: an active hypnotic drug group and a placebo group.
	
	Table~\ref{tab:realdata} shows the estimated values of $\Phi$ and their 95\% confidence intervals for each group.
	
	\begin{table}[htbp]
		\centering
		\caption{Estimated $\Phi$ and 95\% confidence intervals for each group}
		\label{tab:realdata}
		\begin{tabular}{lcc}
			\toprule
			Group & Estimated $\Phi$ & 95\% Confidence Interval \\
			\midrule
			Active drug   & $-0.655$ & $(-0.806,\ -0.503)$ \\
			Placebo       & $-0.453$ & $(-0.591,\ -0.316)$ \\
			\bottomrule
		\end{tabular}
	\end{table}
	
	The negative values of $\Phi$ in both groups indicate a general tendency toward improved sleep latency after the intervention. Although the point estimate for the active drug group ($-0.655$) suggests a greater improvement compared to the placebo group ($-0.453$), the 95\% confidence intervals overlapped, indicating that the difference was not statistically significant.

	These findings demonstrate the applicability of the proposed measure in real clinical data. In particular, the ability to quantify directional changes within each group and to compare the magnitude of such changes across groups provides valuable insights in studies involving ordinal outcomes derived from discretized continuous variables such as sleep latency.

	\section{Conclusion}
	In this study, we proposed a novel measure $\Phi$ to quantify deviations from the Marginal Homogeneity (MH) model in square contingency tables with ordinal categories. Unlike existing measures, the proposed measure captures not only the magnitude but also the direction of change between pre- and post-intervention conditions. 
	he measure is constructed based on the hazard odds ratio of the marginal distributions in discrete time under the MCORmodel, and serves as a directional measure that distinguishes the nature of deviations from the MH model in terms of discrete-time hazard.

	We developed an asymptotic confidence interval for $\Phi$ using the delta method, allowing for inference regarding the direction and strength of marginal changes. Simulation studies demonstrated that the proposed confidence interval provides adequate coverage probabilities in moderate to large sample sizes across a range of scenarios. Additionally, real data analysis using a randomized controlled trial on insomnia treatment showed the practical applicability of the proposed method, effectively capturing improvement trends in sleep latency within each group and highlighting the absence of a statistically significant difference between groups.
	
	The proposed framework offers a flexible and interpretable approach for analyzing ordinal categorical outcomes frequently encountered in clinical and epidemiological studies. In particular, it is well suited to modern database research, where data are often aggregated and variables that were originally measured on continuous scales must be analyzed as ordinal categories. Such situations are becoming increasingly common due to the use of secondary data sources, such as electronic health records and administrative claims databases, where detailed continuous measurements may not be available.
	
	Moreover, the proposed measure may have potential applications in the evaluation of predictive models, particularly those designed to forecast event occurrence times in discrete time intervals. For example, recent studies such as \cite{horn2022} employed deep learning models to estimate period-specific survival probabilities of ICU patients. In such contexts, comparing predicted and observed discrete-time survival patterns using a directional marginal measure like $\Phi$ could offer an informative summary of model performance, especially in terms of whether the model tends to overestimate or underestimate risk over time.
	
	As a future direction, it is worth noting that the proposed measure corresponds to the MCOR model, which captures the difference in discrete-time hazards between the row and column variables. However, in clinical research, hazard ratios rather than absolute hazard differences—are often used to interpret treatment effects. Developing a model that focuses on marginal hazard ratios and constructing a corresponding measure would enhance the clinical interpretability of the results. Such an extension could provide a more intuitive framework for evaluating treatment effects in ordinal categorical data settings, aligning more closely with common practice in time-to-event analyses.


\end{document}